\documentclass[conference]{IEEEtran}
\IEEEoverridecommandlockouts
\usepackage{cite}
\usepackage{amsmath,amssymb,amsfonts}
\usepackage{algorithmic}
\usepackage{graphicx}
\usepackage{textcomp}
\usepackage{xcolor}
\usepackage{multirow}

\usepackage[a4paper, total={184mm,239mm}]{geometry}
\def\BibTeX{{\rm B\kern-.05em{\sc i\kern-.025em b}\kern-.08em
    T\kern-.1667em\lower.7ex\hbox{E}\kern-.125emX}}
\begin{document}

\newcommand{\rohan}[1]{\textcolor{blue}{#1}}%
\setlength{\textfloatsep}{0.25cm}

\title{NOVA: \underline{No}C-based \underline{V}ector Unit for Mapping \underline{A}ttention Layers on a CNN Accelerator}

\author{\IEEEauthorblockN{Mohit Upadhyay, Rohan Juneja, Weng-Fai Wong and Li-Shiuan Peh}
\IEEEauthorblockA{{School of Computing, National University of Singapore} \\
\textit{mohitu@u.nus.edu, rohan@comp.nus.edu.sg, wongwf@nus.edu.sg, peh@nus.edu.sg}}
}

\maketitle


\begin{abstract}

Attention mechanisms are becoming increasingly popular, being used in neural network models in multiple domains such as natural language processing (NLP) and vision applications, especially at the edge. However, attention layers are difficult to map onto existing neuro accelerators since they have a much higher density of non-linear operations, which lead to inefficient utilization of today's vector units. This work introduces NOVA, a NoC-based Vector Unit that can perform non-linear operations {\em within} the NoC of the accelerators, and can be overlaid onto existing neuro accelerators to map attention layers at the edge. Our results show that the NOVA architecture is up to $37.8\times$ more power-efficient than state-of-the-art hardware approximators when running existing attention-based neural networks.


\end{abstract}


%
%

\vspace{-0.75em}
\section{Introduction}
\label{sec:intro}


\vspace{-0.5em}
Attention-based models such as BERT and GPT have recently achieved significant breakthroughs in NLP-based applications over models based on CNN and {\em recurrent neural network} (RNN) based models. BERT has beaten human performance on challenging sentence classification tasks, and ChatGPT is now the talk of the town. 
The attention mechanism is the backbone operation in transformer models: It is a content-based similarity search with a very high density of matrix multiplication and non-linear operations such as ReLU, Softmax and GeLU. These reduce the acceleration benefits of tensor operation acceleration since the non-linear operations are performed by vector units or sent off-chip, creating a performance bottleneck. Non-linear operations can consume up to nearly 40\% of the runtime in models with significant attention layers~\cite{nn-lut}\cite{softermax}. 

There have been multiple hardware accelerators~\cite{ibert}\cite{softermax} proposed for accelerating non-linear operations using hardware-friendly techniques, but typically model accuracy suffers because of the lower precision and approximation techniques. Other approximation approaches for frequently used activation functions in deep neural networks~\cite{approx,efficient-softmax}, such as the use of piecewise linear functions~\cite{online-approx,piecewise-lin} also cause accuracy loss, and are not amenable to efficient hardware implementation. NN-LUT\cite{nn-lut} shows that the accuracy loss can be mitigated by approximating the non-linear functions using an MLP model. This allows for mapping non-linear functions in a hardware-friendly manner. NN-LUT maps the MLP onto LUTs. For many other NoC-based accelerators~\cite{tpu_v4,eyeriss}, adding LUTs for approximating non-linear operations will incur high overhead.


In this paper, we propose NOVA, a NoC-based vector unit that approximates the non-linear operations {\em within} the NoC, relying on the NoC to calculate and broadcast values to different processing elements on-chip. Specifically, the NOVA architecture can be overlaid on top of any NoC-based CNN accelerators to efficiently accelerate attention-based models, at low area and power overheads. 

The key contributions and results of this paper are as follows:
\textbf{(i)} This work introduces NOVA, a NoC-based vector unit that performs on-chip, non-linear approximation by using a neural network to compute non-linear activation functions such as SoftMax, GeLU with much higher area and power efficiency.
\textbf{(ii)} This work is able to map such non-linear operations on the NOVA architecture on top of existing hardware accelerators by using NOVA as a NoC overlay. 
The overall architecture allows mapping transformer architectures onto existing hardware architecture with minimal increase in hardware overhead.
\textbf{(iii)} 
We integrate NOVA with several existing accelerator designs to show that it can be overlaid atop existing systolic array architectures as well as coarse-grained accelerators. We show NOVA is more area and power efficient than existing vector units on average by $3.23\times$ and $16.56\times$ respectively. This also leads to better energy savings than existing state-of-the-art hardware approximators by more than $9.4\times$ for attention based neural network models on average.

Section 2 discusses approximation of non-linear operations in attention-based models and walks through how LUT-based accelerators work. Section 3 details the NOVA hardware architecture and 
Section 4 presents the associated software mapping. Section 5 evaluates NOVA against state-of-the-art hardware approximators. Section 6 covers related works 
while Section 7 concludes the paper.

\vspace{-0.5em}

%
%

\section{Background and Motivation}

\begin{figure}[tbh!]
    \centering
    \includegraphics[height=4cm]{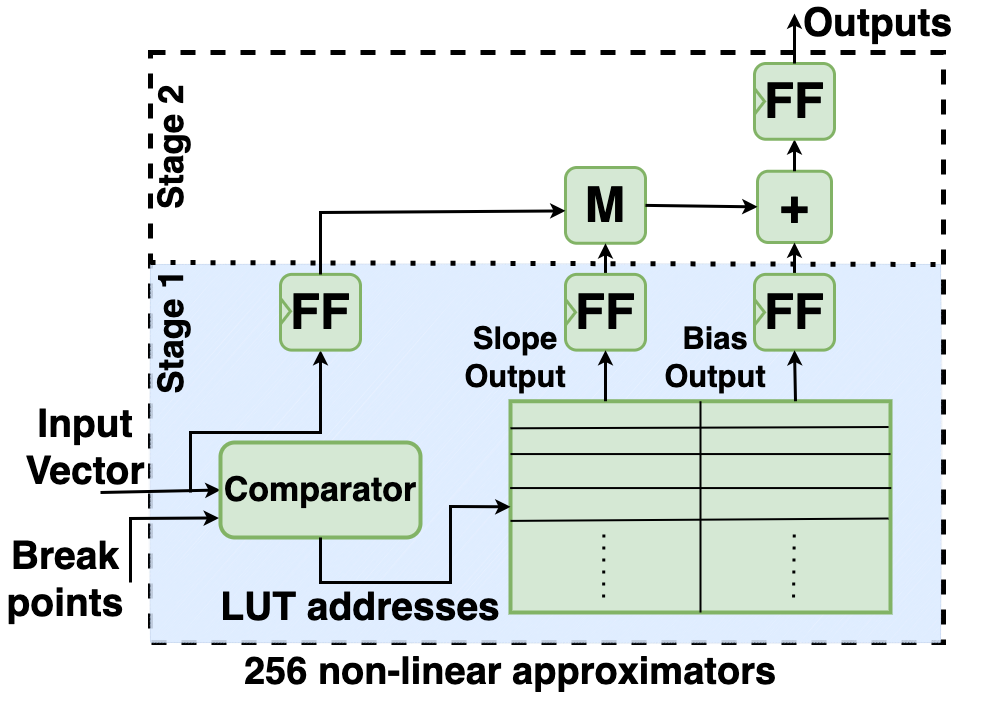}
    \caption{LUT-based approximator (shared by 256 neurons)}
    \label{fig:nn-lut}
    \vspace{-1em}
\end{figure}

\vspace{-0.5em}
{\bf MLP as approximator:} NN-LUT\cite{nn-lut} proposes that a piecewise linear function is used as a universal approximator for non-linear operations where the slope and bias values for piecewise linear operations are learnt by a Multi-Layer Perceptron (MLP). This was used in conjunction with a flexible NPU\cite{samsung_npu} architecture, proposed by Samsung for mobile-based SoCs (System-on-Chips) which comprises of an inner product engine for leveraging the input feature map sparsity. Specifically, it uses a single hardware pipeline with LUT-based architecture with MACs to approximate different non-linear operations, enabling low hardware overheads at minimal accuracy loss (see Fig~\ref{fig:nn-lut}). It leverages the LUTs already present in the Samsung NPU to realize approximation at low overhead.

\begin{figure}[tbh!]
    \centering
    \includegraphics[width=.8\columnwidth, height=4.5cm]{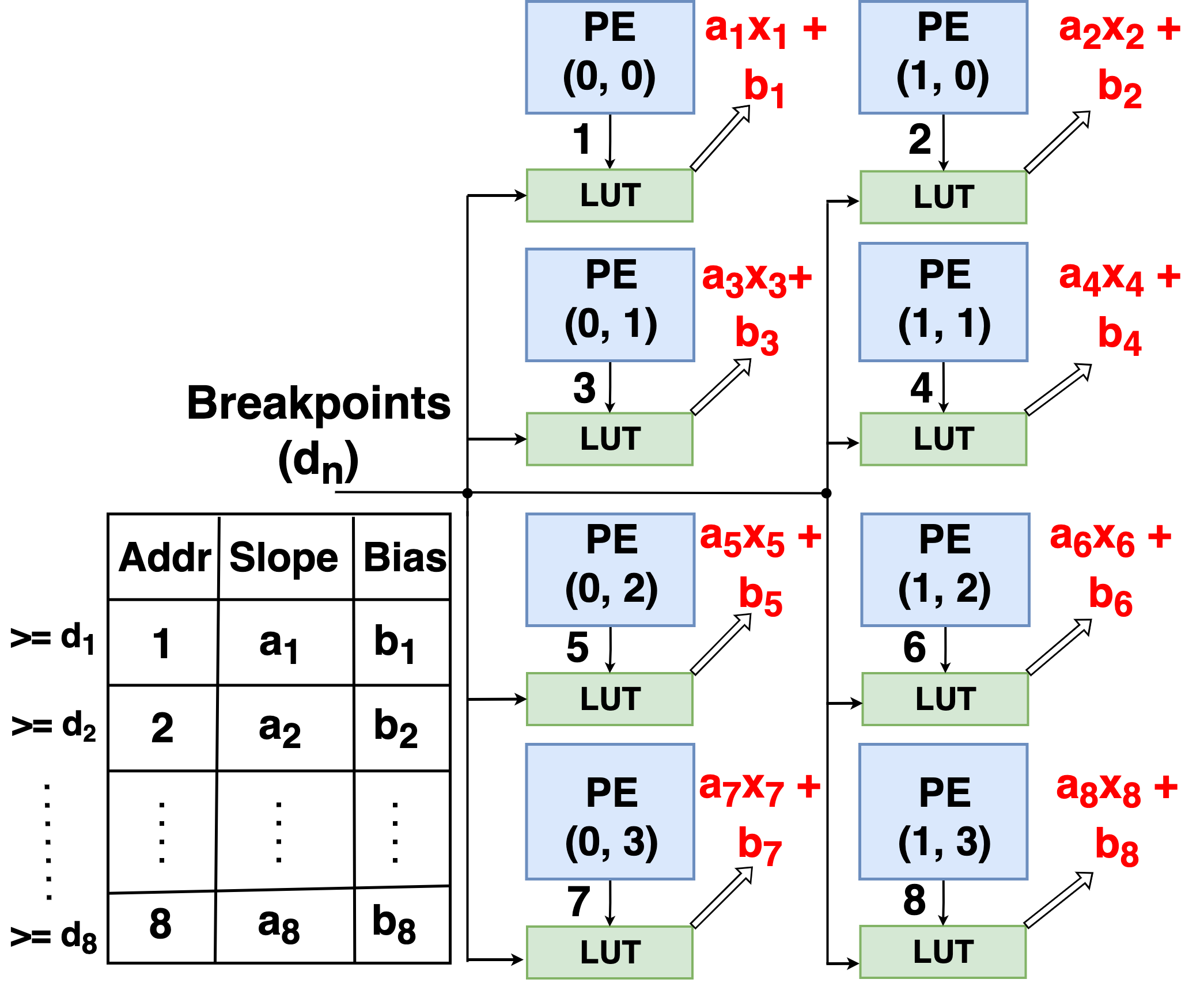}
    \caption{Walkthrough of approximation with LUT-based baseline}
    \label{fig:lut-walkthrough}
\end{figure}

{\bf Walkthrough:} We take a simple walkthrough to understand the inner workings of the NN-LUT architecture\cite{nn-lut} working in conjunction with a spatial accelerator with 8 processing elements (PEs) arranged as a $4\times2$ grid as shown in Fig~\ref{fig:lut-walkthrough}.
In this example, we assume that we use 8 breakpoints to approximate non-linear functions. For simplicity, we assume that each PE has a single output neuron linked to an NN-LUT unit, denoted as $x_1,..,x_8$ such that the outputs fall within different sections 1-8 of the piecewise linear function. Hence, the lookup addresses generated by the comparators connected to the PEs (0,0)-(1,3) are 1 to 8. Hence, the final approximators' output would be $a_1x_1+b_1,..,a_8x_8+b_8$.
The lookup addresses are generated by comparing the input values to the breakpoints ($d_n$). Every cycle, different slope and bias values are fetched based on the lookup addresses (comparator outputs) at every PE. So, the entire non-linear computation takes 2 clock cycles, one for fetching slope and bias values from the LUT and another to perform the MAC operation to get the approximated result.



\vspace{-1em}


{\bf Approximation accuracy:} To verify the effectiveness of an MLP as a non-linear approximator~\cite{nn-lut}, we simulate the accuracy for different CNN, Multi-layer Perceptron (MLP) and Attention benchmarks with MLP-approximated non-linear activation functions, without any retraining on the respective datasets. 
Table~\ref{table:accuracy} shows negligible accuracy loss for different models used with approximated non-linear functions using 16 breakpoints for the approximation as they are sufficient for the commonly used non-linear functions\cite{nn-lut}.


\begin{table}[h!] 
\centering
\resizebox{.7\columnwidth}{!}{
\begin{tabular}{|l|c|c|}
\hline
Neural Network          & Accuracy with & Accuracy with \\
Models                  & Softmax       & Approx. Softmax \\ \hline
MLP (MNIST)             & {97.31}       & {97.31}\\ \hline
CNN (CIFAR-10)          & {63.44}       & {63.44}\\ \hline
MobileNet v1 (CIFAR-10) & {68.56}       & {68.56}\\ \hline
VGG-16 (CIFAR-10)       & {88.30}       & {88.30}\\ \hline
MobileBERT (SQUAD)      & {89.30}       & {89.30}\\ \hline
RoBERTa (SST-2)         & {94.60}       & {94.40}\\ \hline
\end{tabular}
}
\caption{Post approximation accuracy comparison (All models use 16 breakpoints except CIFAR-10 which uses 8)}
\label{table:accuracy}
\vspace{-2em}
\end{table}


{\bf NoC-based accelerators:} Unlike the Samsung NPU, most AI accelerators do not have LUTs and are NoC-based such as the TPU\cite{tpu_v4} and Eyeriss\cite{eyeriss}. These accelerators have a fine-grained architecture where the {\em multiply-and-accumulate} (MAC) operations are mapped onto multiple processing elements (PEs) computing tensor operations, interconnected by NoCs for scalability and energy efficiency. 
Adding an NN-LUT unit to these accelerators will incur additional overhead. Instead, we propose a novel microarchitecture which converts this piecewise mapping function into an in-network routing-based approach. The proposed architecture can be efficiently overlaid 
onto the existing NoCs in NoC-based accelerators.

\vspace{-0.75em}
\label{sec:background}

%
%

\section{NOVA Hardware Architecture}


\subsection{NoC-based approximator hardware architecture}


As previously explained, NN-LUT\cite{nn-lut} uses a MLP to set the breakpoints within the piecewise linear function. The input vector to NN-LUT is compared to the breakpoints to generate lookup addresses to index into LUTs to fetch the slope and bias values. In NOVA, we transform this LUT-based operation into a NoC-based one: NOVA broadcasts the different values of slope and bias across the NoC for matching with the lookup addresses. In simpler terms, the slope and bias values are "stored" in the NoC wires instead of being stored in memories (LUTs). The NoC is arranged in a \textit{line topology} which routes the packets (slope and bias values) in a pre-defined route snaking through the entire NoC, one PE after the other. The NOVA NoC relies on clockless repeaters enabling single-cycle multi-hop transmission, like SMART NoCs~\cite{smart}, to realize fast broadcast across the chip.


We briefly outline the working of the NOVA architecture with 8 PEs in Fig~\ref{fig:nl-noc_routing}. All other parameters are assumed to be the same as that of the walkthrough example of the LUT-based architecture in Section~\ref{sec:background}.

NOVA broadcasts 8 pairs of slope and bias values used to approximate non-linear operations every cycle. In this example, the slope and bias values for the lookup addresses 1-8 along with the tag bits are routed along the entire NoC in cycle 1. Since the lookup address generated at Core (0,0) is 1, hence the slope and bias values for the lookup address 1 are fetched in Core (0,0). Similarly, slope and bias values for the the lookup addresses 2-8 are fetched at Cores (0,1)-(3,1) respectively. After each core fetches the respective slope and bias values, they are sent to the MAC unit to perform the final approximation operation in the next cycle. Our design choice of broadcasting removes the need to store the slope and bias values at each PE, thereby reducing on-chip data redundancy, while the use of a simple line topology minimizes the complexity of the NoC routers, lowering overheads.


We explain the router microarchitecture of the routers and the routing performed within the NOVA NoC in more detail in the following subsections.

\begin{figure}[tbh!]
    \centering
    \includegraphics[height=6cm]{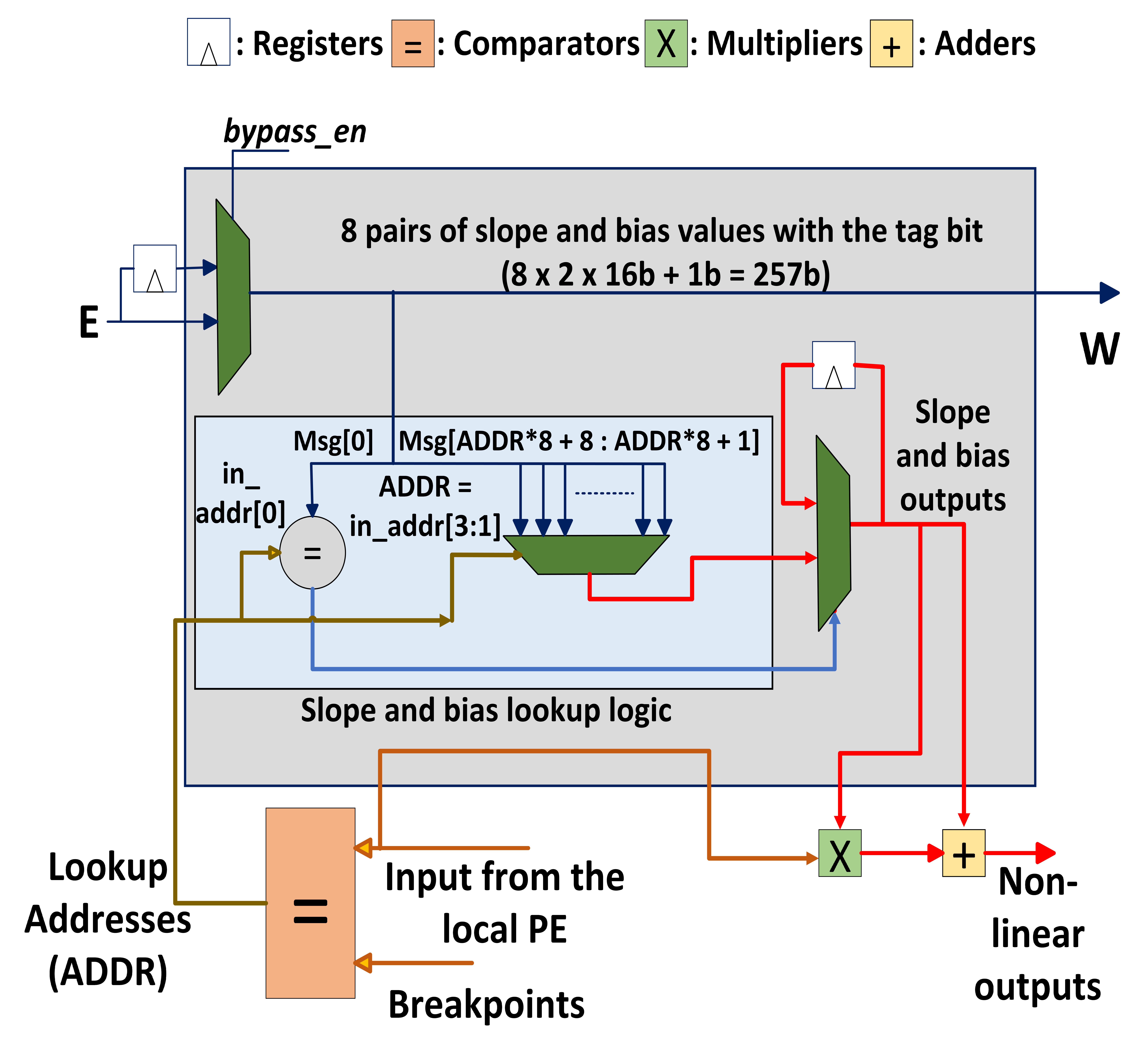}
    \caption{Architecture of NOVA router with the comparator and MAC. Each router has two input and output links, connected in a 1D line topology. Each input and output link is 257 bits wide, encompassing 16 words (8 pairs of slope and bias values) along with their corresponding tag bit. }
    \label{fig:nl-router-microarch}
\end{figure}

\subsubsection{NOVA router micro-architecture (Fig~\ref{fig:nl-router-microarch})}

Each router consists of two input and output ports. The input ports are an east input port from the neighboring router and a local input port from the comparators at each PE. The two output ports are a west output port to the neighboring router and a local output port which returns the approximated non-linear values.

The outputs from each PE are processed by the comparators to generate lookup addresses, which are then sent to the corresponding NOVA router. The NOVA NoC efficiently routes eight pairs of slope and bias values used for the approximation, along with their respective tag bit every cycle across all the PEs. Within each NOVA router, the LSB of each lookup address is used to match against the tag bit of the incoming packet. The remaining bits are used to retrieve the slope and bias values for the corresponding lookup addresses generated from the PE outputs. The PE outputs are multiplied by the respective slope values and the result is added to the bias values to generate the final approximated results.





\vspace{-1em}


\subsubsection{NOVA NoC routing}

\begin{figure}[tbh!]
    \centering
    \includegraphics[height=6cm]{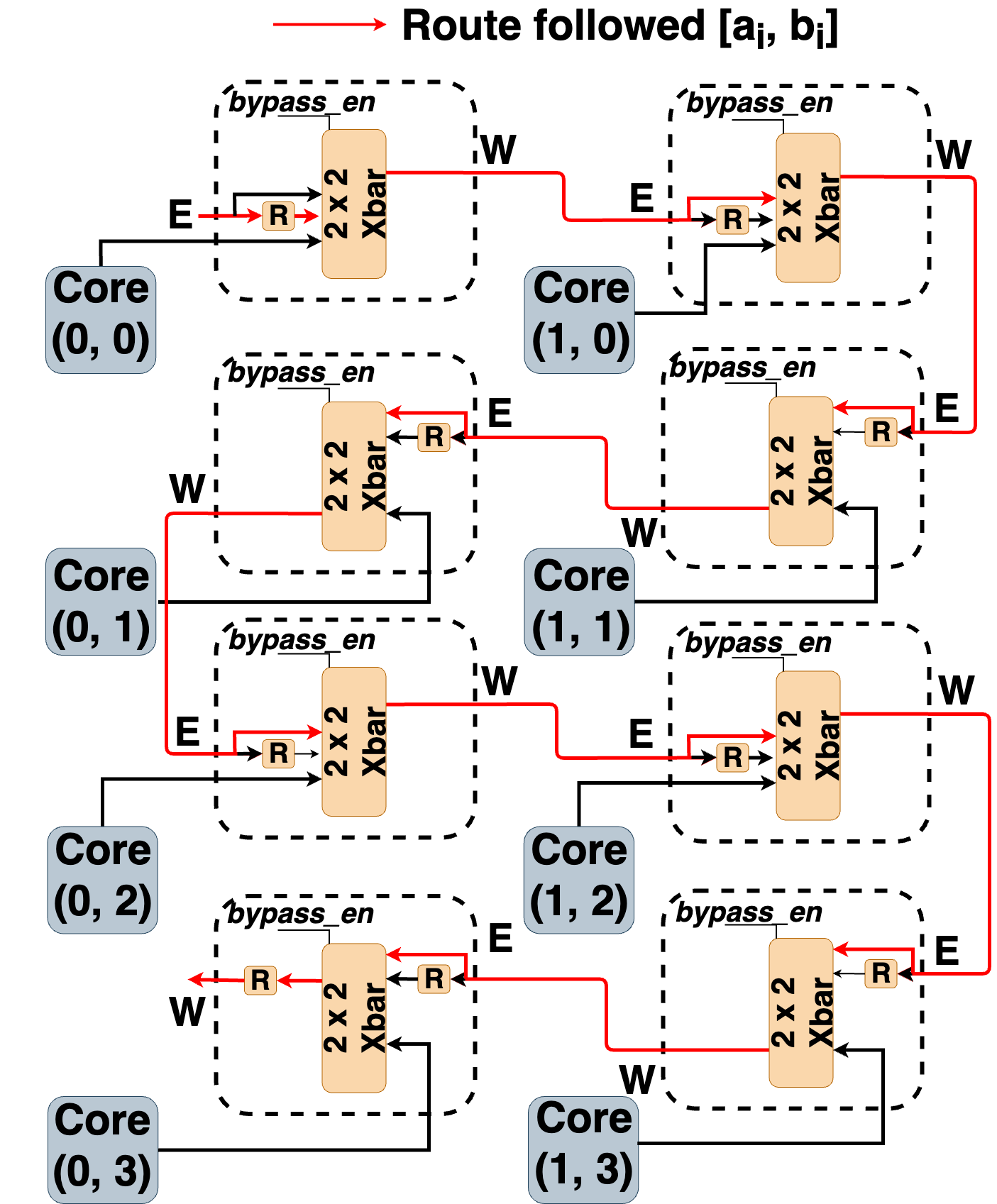}
    \caption{Walkthrough of approximation using NOVA NoC} 
    \label{fig:nl-noc_routing}
    \vspace{-1em}
\end{figure}

Fig~\ref{fig:nl-noc_routing} illustrates a simple example of NOVA's routing, where a message containing the slope and bias values along with a tag bit is routed from Core (0,0) to (0,3). For the sake of simplicity, we illustrate only the router input and output ports from the neighboring routers, excluding the eject ports as they do not contribute to the routing process in the NoC. The NOVA NoC routes the data packets across all the interconnected routers that are linked to their respective cores. Each router's east input port consists of registers (for 8 pairs of slope and bias values) along with a bypass path. There are asynchronous repeaters connected to the output ports which allows NOVA to traverse the NoC in one clock cycle. Since the NoC is organized in a line topology, the route followed by the data packets is fixed which reduces the need for a complex flow control logic and instead just requires setting the router to buffer or forward the data at the input ports as shown in Fig~\ref{fig:nl-noc_routing}.

\vspace{-0.5em}
\subsection{Integrating NOVA NoC with third-party accelerators}

\vspace{-0.5em}


\begin{figure*}[h!]
    \centering
    \includegraphics[height=6cm]{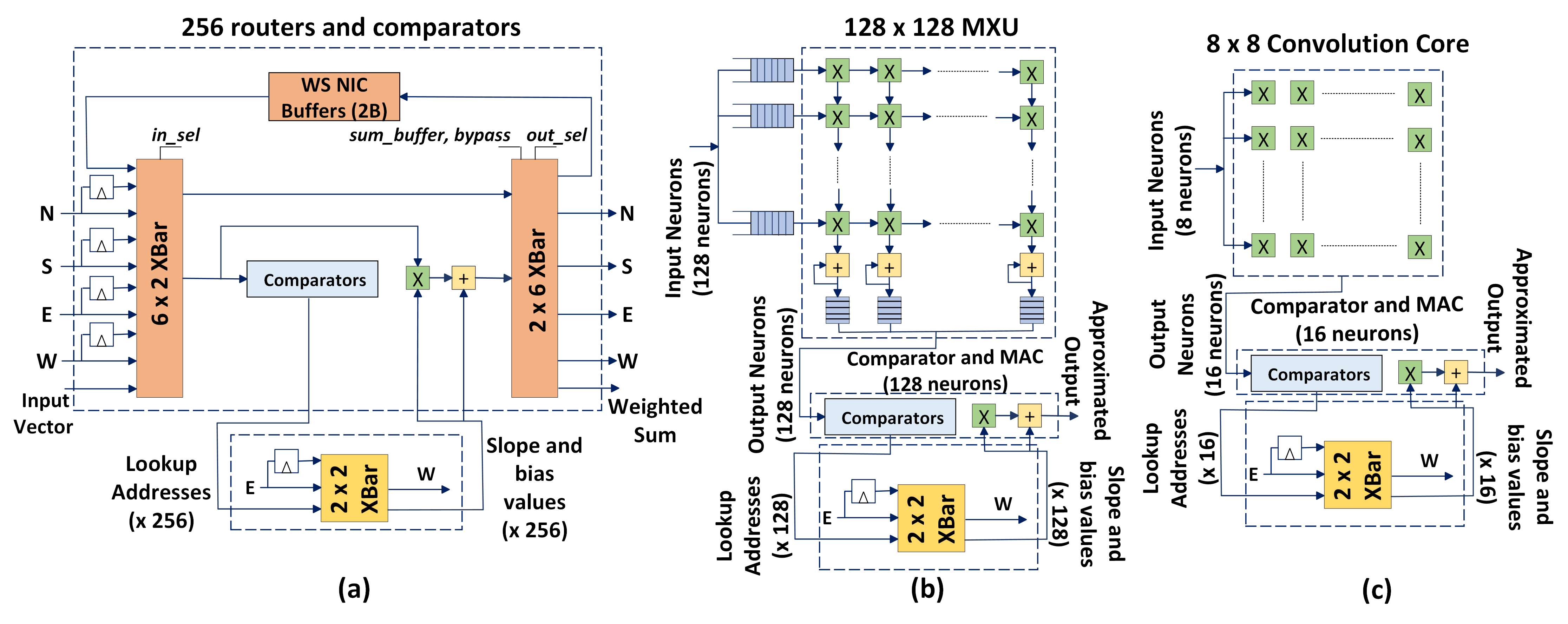}
    \caption{Integrating NOVA with (a) REACT WS routers, (b) TPU v3/v4 MXU, (c) NVDLA Convolution Core}
    \label{fig:nl-router-accelerator}
    \vspace{-2em}
\end{figure*}

\subsubsection{Integrating NOVA NoC with REACT (Fig~\ref{fig:nl-router-accelerator}(a))}

NOVA NoC is connected to the Weighted Sum (WS) NoC within REACT\cite{react} where the comparators are used to generate the lookup addresses along with the MAC units for the non-linear approximation. The REACT's WS router is altered to change it to $6\times2$ input crossbar which fetches output data from an input port coming from the current or a neighboring PE. One output is used to bypass the NOVA NoC, while another outputs to the comparators for generating the respective lookup addresses, which are sent to the NOVA NoC to fetch the respective slope and bias values for the non-linear approximation. The local outputs from the NOVA approximator are sent to the $2\times6$ output crossbars to send to one of the output ports.

\subsubsection{Integrating NOVA NoC with a Systolic Array Architecture (Fig~\ref{fig:nl-router-accelerator}(b))}

The TPUs' tensorcores~\cite{tpu_v4} consists of matrix multiply units termed MXUs with multiply/accumulators organized in a systolic array. NOVA is connected to the MXUs to provide non-linear approximation. 
Specifically, the output of the MXUs are sent to the comparators to generate the respective lookup addresses, which are fed to the NOVA routers to fetch the respective slope and bias values and sent to the MAC units for approximation.


\subsubsection{Integrating NOVA NoC with NVDLA (Fig~\ref{fig:nl-router-accelerator}(c))}

NVDLA\cite{nvdla} is a modular inference accelerator that can be configured for diverse applications. It comprises  multiple components such as Convolution Engines for high-performance convolution, Single Data Processor (SDP) for computing activation functions, Planar Data Processor for pooling and Processing engines for advanced normalization functions.
Each NOVA router is connected with an NVDLA convolution core\cite{nvdla}.




\vspace{-0.75em}
\label{sec:architecture}

%
%

\section{Mapping of Non-linear Operations on NoC Accelerators with NOVA}

\vspace{-0.5em}

NOVA's NoC uses a piecewise linear function for non-linear approximation where a Multi-Layer Perceptron (MLP) model generates the slope and bias values for the piecewise linear function as shown in NN-LUT\cite{nn-lut}. The MLP model has 2 layers where the number of nodes in the hidden layer represent the number of breakpoints required for non-linear approximation. The MLP model is trained at compile time since the non-linear functions within the model are already known.

The NOVA mapper schedules the cycle-by-cycle operation of NOVA NoC, ensuring correct functionality of the lookup operation across the NoC. It thus needs to set up the NoC to broadcast the slope and bias values appropriately based on the number of breakpoints for non-linear approximation. Since NOVA's NoC broadcasts 8 pairs of slope and bias values in every clock cycle, it takes multiple cycles for the higher number of breakpoints needed for non-linear approximation. In order to keep the lookup latency to 1 cycle,  NOVA's NoC runs at higher clock frequency that is set by the mapper at runtime.

\vspace{-0.75em}

\label{sec:mapping}
%
%

\section{Evaluation}


\subsection{Experimental methodology}

We evaluate our NOVA NoC overlaid atop several popular deep learning accelerators. Table~\ref{table:arch_parameters} lists the NOVA NoC architecture (number of cores, number of output neurons at each core, on-chip memory size) for each accelerator.
For REACT, we evaluated it with and without NOVA added to the WS NoC for non-linear approximation. For TPU, we evaluated two configurations of the accelerator modeled after the TPU-v3 and TPU-v4\cite{tpu_v4} configurations where each MXU is a $128\times128$ systolic array. We compare MXU cores paired with the LUT-based vector units vs. MXU cores with a similar number of NOVA routers. 
For NVDLA, a smaller Nvidia Jetson NX configuration\cite{xavier} SoC with NVDLA cores is modeled using the ESP\cite{esp} tool, with its native LUT-based SDP vs. with NoC-based NOVA.


\begin{table}[h!]
\centering
\resizebox{\columnwidth}{!}{
\begin{tabular}{|c|c|l|l|c|}
\hline
\begin{tabular}[c]{@{}c@{}}Hardware \\ Accelerator\end{tabular} & \begin{tabular}[c]{@{}c@{}}Num of \\ NOVA routers\end{tabular} & \begin{tabular}[c]{@{}l@{}}Num of neurons\\ per NOVA router\end{tabular} & \begin{tabular}[c]{@{}l@{}}On-chip \\ memory\end{tabular} & \begin{tabular}[c]{@{}c@{}}Operating  \\ Frequency \\ (0.8V)\\ {[}MHz{]}\end{tabular} \\ \hline
REACT            & 10 & 256 & 768 kB & 240 \\ \hline
TPU v3-like      & 4  & 128 & 42 MB  & 1400 \\ \hline
TPU v4-like      & 8  & 128 & 42 MB  & 1400 \\ \hline
Jetson Xavier NX & 2  & 16  & 256 kB & 1400 \\ \hline
\end{tabular}
}
\caption{Accelerator parameters integrated with NOVA}
\label{table:arch_parameters}
\vspace{-2em}
\end{table}

We modeled the NOVA NoC and baseline LUTs in RTL in order to analyse the trade-offs of using NOVA vs LUTs for non-linear approximation. We used 16 breakpoints, i.e., a maximum of 16 sets of slope and bias values to approximate non-linear operations, as that achieves excellent approximation accuracy (see Table~\ref{table:accuracy}).

The design was implemented and verified on SystemVerilog HDL and its functionality was verified using the Synopsys VCS tool. Synopsys Design Compiler was used to synthesize the RTL design to gate-level netlists on a commercial 22nm CMOS process, with the foundry’s memory-compiled SRAM cells along with wire modeling performed using the Cadence Genus tool.
As NOVA replaces significant number of registers and memory elements with wires, and wiring overhead can be under-estimated by synthesis, we performed placement and routing on NOVA and the baselines in order to accurately model the overheads. This was done with the Cadence Genus Innovus toolchain. The clockless repeaters in NOVA NoC require timing analysis to be carried out across the entire NoC line topology, with the clock edge registered at NoC inputs.
As the 257-bit NOVA NoC, which is shared across all the neurons, can only broadcast 8 pairs of slope and bias values each cycle, the NOVA NoC operates at a clock frequency $2\times$ that of the comparators and the MAC, when used with 16 breakpoints to keep NOVA's overall latency to a single cycle. Running NOVA NoC at $2\times$ the frequency of the base accelerators
can be handled by standard IC design techniques~\cite{cdc}.


{\bf Scalability:} Based on the place and route timing results, a maximum of 10 routers with clockless repeaters placed 1mm apart can be traversed at 1.5 GHz clock. Hence, all the above NOVA configurations with different accelerators having $<=$10 routers can complete a broadcast traversal within a cycle. 
Scaling beyond 10 routers will lead to the traversal taking multiple clock cycles, which trades off the latency benefits with potentially lower clock frequency and power savings.
\vspace{-0.25em}


\subsection{Synthesis results: Baseline LUT-based Vector Unit vs. NOVA}

\begin{figure}[tb!]
    \centering
    \includegraphics[width=\columnwidth]{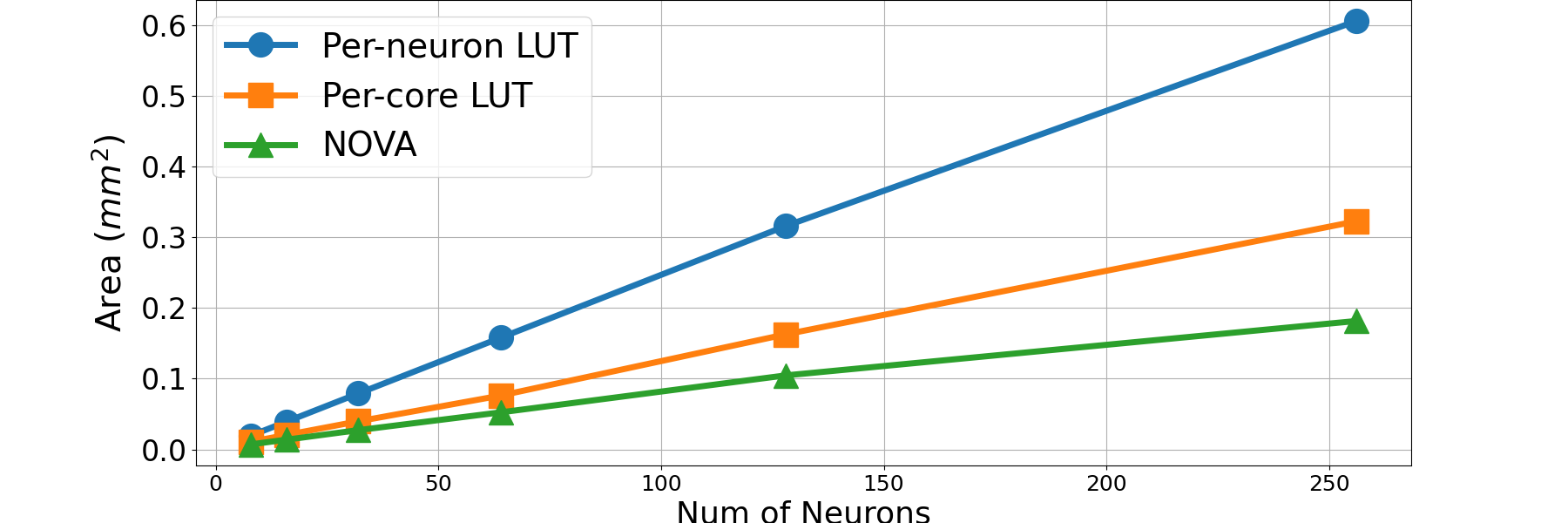}
    \caption{NOVA router area vs no. of neurons mapped per router}
    \label{fig:nl-router-area}
\end{figure}

\begin{figure}[tb!]
    \centering
    \includegraphics[width=\columnwidth]{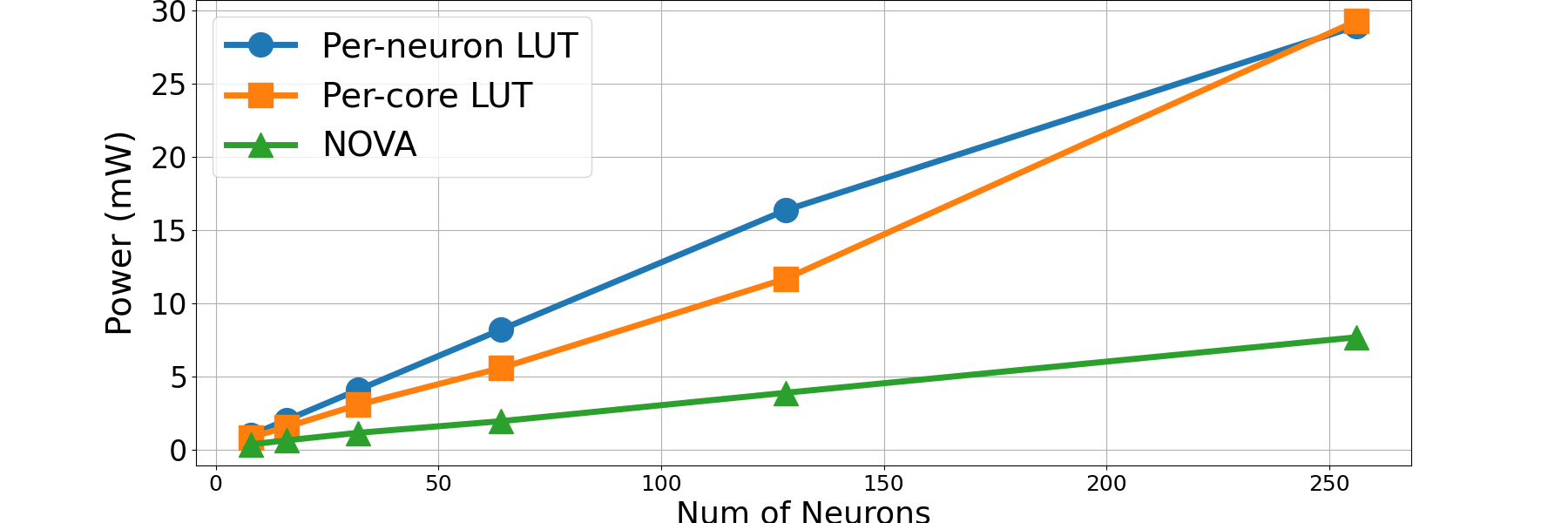}
    \caption{NOVA router power vs no. of neurons mapped per router}
    \label{fig:nl-router-power}
\end{figure}

\vspace{-0.25em}
We modeled two versions of LUT-based vector units with different accelerators, a per-neuron LUT which maps each LUT (storing the slope and bias values) to every neuron which uses single ported banks and the second version, a per-core LUT  which maps all the neurons to one multi-ported LUT bank, which reduces the need to store multiple copies of the same data within a core to reduce the redundancy. These two versions give an estimate of two extreme variations of LUT-based architectures. The size of each LUT bank is kept at 64 bytes each since 16 pairs of the slope and bias values are stored in each LUT. Both baseline LUT versions operate at the same clock frequency as the rest of the accelerator, so NOVA's latency is identical to that of the baseline. As shown in Figs~\ref{fig:nl-router-area} and~\ref{fig:nl-router-power}, NOVA is much more power and area-efficient than the baselines and scales better with neuron count.

\subsection{Synthesis Results: REACT integrated with NOVA}


\subsubsection{Area}



Having LUT-based non-linear approximators within REACT leads to an additional area overhead of around 31\% and 19.2\% for the two versions of the LUT-based vector units respectively on top of on-chip REACT die area. On the other hand, having a NOVA approximator integrated with REACT architecture adds just 9.11\% on top of on-chip REACT die area.
Comparing the NOVA NoC based approximator to the two versions of LUT-based approximate hardware, it has an overall area savings of $3.34\times$ and $1.78\times$ respectively.

\subsubsection{Power}

The LUT-based approximators when integrated with REACT has a power overhead of 289.08mW and 292.57mW respectively. The per-core LUT has higher number of ports to facilitate the sharing of each LUT output across all neurons, which leads to higher power consumption than the per-neuron LUT baseline. The NOVA NoC-based non-linear approximator when operating at a higher clock frequency has an additional power overhead of only 117.51mW for REACT. Thus, when comparing the NOVA NoC approximator with the two different versions of the LUT-based approximators, there is a power saving of $2.5\times$ on average.

\vspace{-0.5em}
\subsection{Synthesis Results: TPU v3/v4 integrated with NOVA}

\vspace{-0.5em}

\subsubsection{Area}
Each NOVA NL router and each LUT-based vector unit (per-core and per-neuron LUT-based vector unit) is shared among 128 neurons since each MXU in the TPU configurations has a $128\times128$ multiplier array. The area overhead for NOVA's NoC for a TPU-v3 like and TPU-v4 like configuration is 0.41$mm^2$ and 0.82$mm^2$ respectively. In comparison, the area overhead for the LUT-based vector units for the TPU-v3 like configuration is 1.267$mm^2$ and 1.004$mm^2$ respectively. Similarly for a TPU-v4 like configuration, the area overhead for both the LUT-based vector units is 2.534$mm^2$ and 2.008$mm^2$. NOVA's NoC has an area improvement of over $3\times$ in comparison to the LUT-based vector units.

\subsubsection{Power}

Having LUT-based non-linear approximators when integrated with TPUv4-like architecture has an additional power overhead on average of around 764.5mW and 1.724W for the two different configurations respectively. 
Though the NOVA NoC operates at a maximum of $2\times$ higher clock frequency, it still only take up 184.83mW, leading to a power savings of over $9.4\times$ vs. the baseline LUT-based vector units.

\subsection{Synthesis Results: NVDLA integrated with NOVA}

\vspace{-0.5em}
In order to evaluate NVDLA with NOVA, we synthesized the SoC with the configurations as mentioned in Section III.D.

\subsubsection{Area}

Each NOVA router/LUT is shared among 16 neurons since each NVDLA core has 16 output neurons. The area overhead for NOVA's NoC for an NVDLA core is 0.0276$mm^2$. In comparison, the area overhead for the NVDLA's SDP is 0.1382$mm^2$.
We observe that NOVA's NoC has an area improvement of over $4.99\times$ in comparison to the existing NVDLA's SDP engine which is LUT-based.

\subsubsection{Power}

NVDLA's SDP has a power overhead of 48.867mW. Even though the NOVA NoC operates at a maximum of $2\times$ higher clock frequency, still in contrast it has an additional power overhead of only 1.294mW. It can be seen that the NOVA NoC has power savings of over $37.8\times$.

Table~\ref{table:nl_noc_acc} shows overhead of the NVDLA's SDP in comparison to the NOVA NoC when integrated with the NVDLA cores. For the NVDLA-based SoC, the SDP consumed around 1.2\% of the entire SoC power, whereas in comparison NOVA consumes $37.8\times$ less power.


\begin{table}[h!]
\centering
\resizebox{\columnwidth}{!}{
\begin{tabular}{|c|c|c|c|}
\hline
Accelerator & Hardware Approximator & Area ($mm^2$) & Power (mW) \\ \hline
\multirow{3}{*}{REACT} & naive LUT (per-neuron LUT) & 6.058 & 289.08 \\ \cline{2-4} & naive LUT (per-core LUT) & 3.226 & 292.57 \\ \cline{2-4} & NOVA NoC & 1.817 & 117.51 \\ \hline
\multirow{3}{*}{TPU v3-like} & naive LUT (per-neuron LUT) & 1.267 & 382.468 \\ \cline{2-4} & naive LUT (per-core LUT) & 1.004 & 862.472 \\ \cline{2-4} & NOVA NoC & 0.414 & 103.78 \\ \hline
\multirow{3}{*}{TPU v4-like} & naive LUT (per-neuron LUT) & 2.534 & 764.936 \\ \cline{2-4} & naive LUT (per-core LUT) & 2.008 & 1724.94 \\ \cline{2-4} & NOVA NoC & 0.82 & 184.83 \\ \hline
\multirow{2}{*}{\begin{tabular}[c]{@{}c@{}}Jetson\\ Xavier\end{tabular}} & NVDLA SDP & 0.1382 & 48.867 \\ \cline{2-4} & NOVA NoC & 0.0276 & 1.294 \\ \hline
\end{tabular}
}
\caption{Hardware overhead of NOVA versus different LUT-based approximators (on top of existing accelerators)}
\label{table:nl_noc_acc}
\end{table}

\vspace{-2.5em}
\subsection{Energy Evaluation Results}


We ran five attention benchmarks namely MobileBERT-base, MobileBERT-tiny\cite{mobilebert}, RoBERTa\cite{roberta}, BERT-tiny and BERT-mini\cite{bert} which are representative of real-world NLP based tasks. These are run in conjunction with the SCALE-Sim\cite{scalesim} toolchain to simulate the NOVA's performance atop the TPU v3-like and v4-like architectures. The energy consumption numbers are calculated using the respective power consumption number from the synthesis results.



\begin{figure}[tb!]
    \centering
\includegraphics[width=\columnwidth]{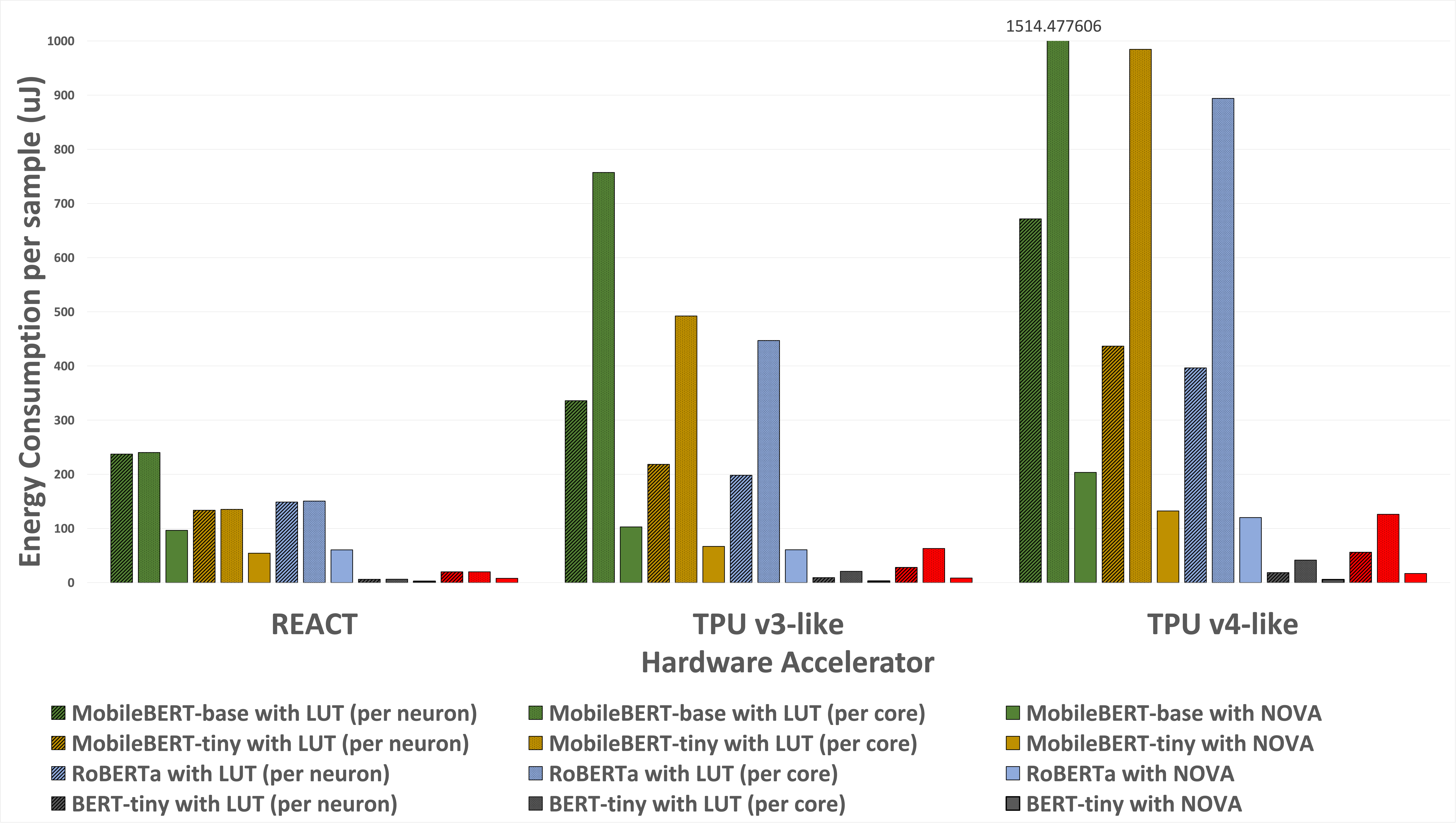}
    \caption{Energy consumption overhead for different non-linear approximator hardware for BERT-like applications}
    \label{fig:energy}
\end{figure}

We evaluate the energy consumption for NOVA and baseline LUT-based architectures atop the REACT, TPU v3 and TPU v4-like accelerators (Table~\ref{table:arch_parameters}). 
The NVDLA SDP energy numbers were not modeled due to insufficient details to accurately model the energy consumption. 
For our evaluations, we use a sequence length of 1024 for all the accelerator configurations in Fig~\ref{fig:energy} except REACT where the sequence length is kept at 128 for the evaluations, since REACT is targeted for the edge, where smaller sequence lengths are more representative. Fig~\ref{fig:energy} shows the energy consumption per inference sample by using NOVA in comparison to using LUT-based approximators with given accelerators for the BERT-based workloads. As shown, overall energy overhead for a LUT-based hardware approximator is as high as $7.5\times$ the energy consumption for NOVA architecture for systolic array based configurations.
As the number of approximation queries typically corresponds to the number of sentences used in NLP applications (BERT-based applications in Fig~\ref{fig:energy}), a higher number of sentences (queries) means higher overall energy consumption. 

Integrating the TPU-v4 configuration accelerator with NOVA's NoC enables mapping attention layers that have lower area and power overhead compared to the LUT-based baseline. 
On average, the BERT-based benchmarks running on the TPU-v4 with NOVA have an energy overhead of only 0.5\% while the LUT-based baselines with TPU-v4 has an energy overhead of $9.4\times$ and $4.14\times$ per input sample, respectively.

\label{sec:evaluation}

%
%

\section{Related Works}




{\bf Non-linear Hardware Approximators:} 
NACU\cite{nacu} proposes a dynamically reconfigurable arithmetic unit for multiple non-linear functions (sigmoid, hyperbolic tangent and exponential function) by using separate computation pipelines for the functions. Softermax\cite{softermax} is a hardware-friendly softmax approach that replaces the exponential function base with 2. 

{\bf Approximation methodologies for non-linear operations:} 
SOMALib\cite{somalib} proposes a library of VLSI implementations for commonly used activation functions. 
It uses Cartesian genetic programming to generate optimized gate-level designs of approximate and exact activation functions. 
I-BERT\cite{ibert} proposes a novel quantization methodology to create integer-only hardware pipelines for the commonly used non-linear functions for transformer-based models. However, the use of integer multipliers, adders, shifters and a divider leads to higher overhead in comparison to NN-LUT.


Unlike others, NOVA leverages NN-LUT's~\cite{nn-lut} mapping to approximate non-linear functions, thereby substantially reducing hardware overhead as shown in Table~\ref{table:nova_comp}. 

\begin{table}[h!]
\centering
\resizebox{\columnwidth}{!}{
\begin{tabular}{|c|c|c|c|}
\hline
\begin{tabular}[c]{@{}c@{}}Non-linear\\ approximator\end{tabular} & Tech node & \begin{tabular}[c]{@{}c@{}}Area\\ ($\mu m^2$)\end{tabular} & \begin{tabular}[c]{@{}c@{}}Power\\ (mW)\end{tabular} \\ \hline
NACU\cite{nacu} & 28 nm & 9671 & 2.159 (sigmoid), 1.95 (tanh), 3.74 (exp) \\ \hline
I-BERT\cite{ibert} & 22 nm & 2941 & 0.201 \\ \hline \hline
NOVA & 22 nm & 898.75 & 0.046 \\ \hline
\end{tabular}
}
\caption{Hardware overhead of NOVA vs NACU}
\label{table:nova_comp}
\end{table}





\label{sec:related_works}



%
%

\section{Conclusion}


This work introduces NOVA, a NoC-based approximator for computing non-linear functions. It transforms a lookup table-based approximation into a NoC-based mechanism. The proposed NOVA NoC can be readily used to extend existing accelerators for attention-based models. It demonstrates the promise of using wires in place of LUTs in AI accelerators.
\vspace{-0.75em}
\label{sec:conclusion}

%
%

\section{Acknowledgements}

{\small This work by the author(s) is fully supported by the Advanced Research and Technology Innovation Centre (ARTIC), the National University of Singapore under Grant (A-0008455-00-00)}
\label{sec:acknowloedgements}

\bibliographystyle{IEEEtranS}

\bibliography{refs}


\end{document}